\newcommand\Sul{Sul--Cu$_2$Cl$_4$}
\documentclass[prl,twocolumn,floatfix,preprintnumbers,amsmath,amssymb,superscriptaddress]{revtex4}

\usepackage{graphicx}% Include figure files
\usepackage{dcolumn}% Align table columns on decimal point
\usepackage{bm}% bold math
\begin{document}

\title{Excitations in a four-leg antiferromagnetic Heisenberg spin tube}

\author{V. O. Garlea}
 \email{garleao@ornl.gov}
\affiliation{Neutron Scattering Sciences Division, Oak Ridge National Laboratory, Oak Ridge, Tennessee 37831, USA.}
\author{A. Zheludev}
\affiliation{Neutron Scattering Sciences Division, Oak Ridge National Laboratory, Oak Ridge, Tennessee 37831, USA.}
\author{L.-P.~Regnault}
\affiliation{CEA-Grenoble, DRFMC-SPSMS-MDN, 17 rue des Martyrs, 38054 Grenoble Cedex 9, France.}
\author{J.-H.~Chung}
\altaffiliation{Currently at:~Department of Physics, Korea University, Seoul 136-701, Korea.}
\affiliation{ NCNR, National Institute of Standards and Technology, Gaithersburg, Maryland 20899, and University of Maryland, College Park, Maryland, 20742, USA.}
\author{Y. Qiu}
\affiliation{ NCNR, National Institute of Standards and Technology, Gaithersburg, Maryland 20899, and University of Maryland, College Park, Maryland, 20742, USA.}
\author{M. Boehm}
\affiliation{Institut Laue Langevin, 6 rue J. Horowitz, 38042 Grenoble Cedex 9, France.}
\author{K. Habicht}
\affiliation{BENSC, Hahn-Meitner Institut, D-14109 Berlin, Germany.}
\author{M. Meissner}
\affiliation{BENSC, Hahn-Meitner Institut, D-14109 Berlin,
Germany.}
\date{\today}

\begin{abstract}
Inelastic neutron scattering is used to investigate magnetic
excitations in the quasi-one-dimensional quantum spin-liquid
system Cu$_2$Cl$_{4}\cdot$ D$_8$C$_4$SO$_2$. Contrary to
previously conjectured models that relied on bond-alternating
nearest neighbor interactions in the spin chains, the dominant
interactions are actually next-nearest-neighbor in-chain
antiferromagnetic couplings. The appropriate Heisenberg
Hamiltonian is equivalent to that of a $S=1/2$ 4-leg spin-tube
with almost perfect one dimensionality and no bond alternation. A
partial geometric frustration of rung interactions induces a small
incommensurability of short-range spin correlations.
\end{abstract}

\pacs{75.10.Jm, 75.50.Ee, 78.70.Nx}

\maketitle

The $S=1/2$ antiferromagnetic (AF) Heisenberg spin ladder with an
even number of legs is the cornerstone model in low-dimensional
quantum magnetism~\cite{Rice93,Barnes93,Gopalan94,Barnes94}. Its
magnetic long range order is totally destroyed by zero-point
quantum fluctuations, yet the ``spin liquid'' ground state is not
entirely structureless, due to hidden topological non-local
``string'' correlations~\cite{denNijs,Kennedy}. Extensive
theoretical studies lead to a profound understanding of spin
ladders and related models, including the origin of the energy
gap~\cite{Rice93,Barnes93,Gopalan94}, multi-particle continua and
bound states~\cite{Barnes94,Sushkov98}, and the unique gapless
quantum-critical phase that can be induced by external magnetic
fields~\cite{Giamarchi,Furusaki}. Only a handful of experimental
realizations of the ladder model have been found to date.
Unfortunately, while some of these, like IPA-CuCl$_3$~\cite{Masuda06,Garlea07}, are not perfectly
one-dimensional, others, like (La,Ca,Sr)$_{14}$Cu$_{24}$O$_{41}$~\cite{Matsuda,Notbohm07}, have
energy scales and phase diagrams that can not be accessed in an experiment. The search for new prototype materials continues.

One compound recently discussed in this context is Cu$_2$Cl$_{4}\cdot$ D$_8$C$_4$SO$_2$ (\Sul). It has a conveniently
small spin gap $\Delta\approx 0.5$~meV, and was described as being composed of $S=1/2$ double-chains with possible geometrically frustrated diagonal rung interactions~\cite{Fujisawa03,Fujisawa05}. However, the gap in \Sul\ could be fully accounted for without invoking the rung interactions,~\cite{Fujisawa03} by a slight structural alternation
in the bond lengths. The corresponding bond-alternating $S=1/2$ chain can be viewed as an array of interacting AF spin dimers, like those found in CuGeO$_3$~\cite{Hirota94,Regnault96} and Cu(NO$_3$)${_2}\cdot$2.5 D$_2$O~\cite{Eckert79}. Systems of this type realize a simple ``local'' spin liquid ground state that lacks the hidden symmetry violation or the translational invariance of the ladder model. Below we present the results of inelastic neutron scattering experiments that reveal that \Sul\ is actually {\it not} a spin-dimer material, but neither it is a triangular ladder. Its key magnetic interactions lead to the formation of {\it uniform 4-leg spin tubes} with no bond alternation and almost perfect 1D character. A geometric frustration stabilizes unique {\it incommensurate} dynamic spin correlations.

\begin{figure}[bp]
\includegraphics[width=3.0in]{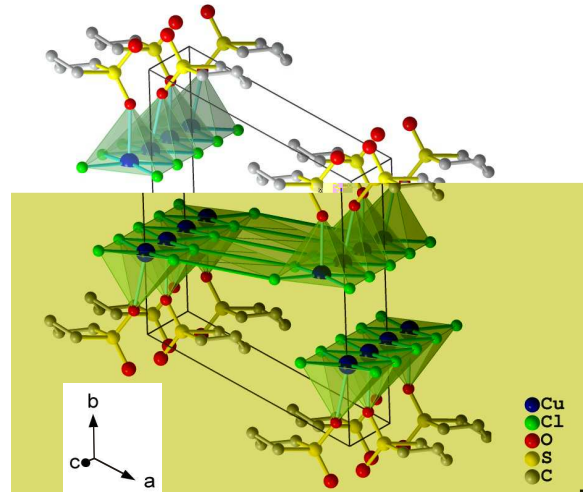}
\caption{(Color online) Schematic view of the crystal structure of \Sul. Dominant AF interactions between $S=1/2$ Cu$^{2+}$ ions are mediated by Cu-Cl-Cl-Cu superexchange bridges.} \label{structure}
\end{figure}

\Sul\ crystallizes in a triclinic space group $P\bar{1}$ with $a =
9.42$ \AA , $b = 10.79$ \AA , $c = 6.62$ \AA , $\alpha =
98.87^{\circ}$, $\beta = 95.25^{\circ}$, and $\gamma =
120.85^{\circ}$~\cite{Fujisawa03}. Pairs of $S=1/2$ chains built
of edge-sharing CuCl$_5$O octahedra run along the crystallographic
$c$-axis (Fig.~\ref{structure}). As mentioned above, due to the
presence of two nonequivalent Cu$^{2+}$ sites, there is a slight
alternation (about 0.5\%) in the distance between consecutive
magnetic ions, Cu$_1$ and Cu$_2$. In the original model of
Ref.~\cite{Fujisawa03}, these chains form the spin ladder legs,
while the rungs involve nearby Cu$^{2+}$ cations packed along the
$b$ axis. Both rung and leg magnetic interactions are established
via Cu-Cl-Cu superexchange pathways. A key feature of this model
is that the distance between consecutive spins on each leg is
$c/2$, so that the 1D AF zone-centers are located at integer
values of $l$, where $(h,k,l)$ are wave vector's reciprocal
coordinates.

\begin{figure}[tbp]
\includegraphics[width=3.1in]{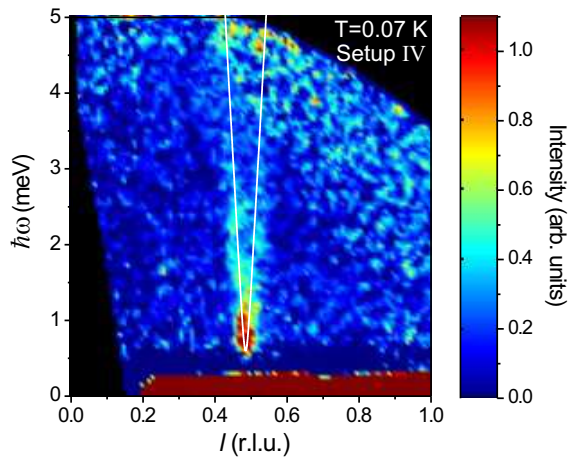}
\caption{Time of flight spectrum measured in \Sul\ in projection
onto the c$^\ast$ axis. The solid line is the
fitted magnon dispersion relation, as described in the text.}
\label{tof}
\end{figure}

Previous neutron scattering studies employed thermal neutrons and
protonated \Sul~ samples~\cite{JRR3}, but failed to detect any
magnetic scattering. In our experiments we instead utilized fully
deuterated \Sul\ single crystals, and put emphasis on cold neutron measurements. Most data were taken using
ten to fifteen single crystals, with total masses of 1.5--2.5~g,
co-aligned to an irregular cumulative mosaic of 3.5$^{\circ}$.
Inelastic data were collected on the SPINS cold-neutron 3-axis
spectrometer at NCNR, with $E_\mathrm{f}=3.7$~meV neutrons, a
cooled BeO filter after the sample, and a horizontal focusing
pyrolitic graphite (PG) analyzer (Setup I). Additional data were
taken using the IN14 spectrometer at ILL, in a similar
configuration (Setup II). Measurements with an enhanced wave
vector resolution were performed on the V2 3-axis spectrometer at
HMI, with a narrow focused PG analyzer and a smaller $\sim 1$~g
sample with a symmetric triangular mosaic spread of $1.9^\circ$
(Setup III). Time-of-flight (TOF) experiments were performed at
the Disc Chopper Spectrometer (DCS) at NCNR using a fixed incident
energy $E_\mathrm{i}=6.67$~meV (Setup IV).

\begin{figure}[tp]
\includegraphics[width=3.3in]{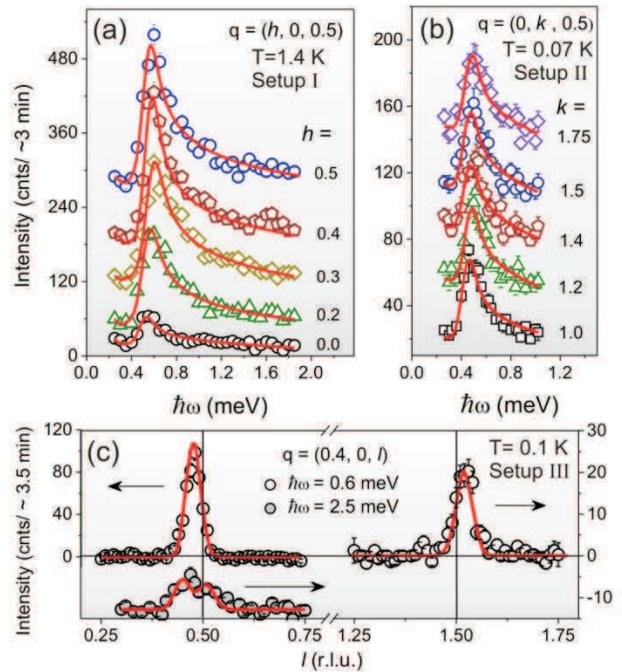}
\caption{(Color online) (a,b) Energy scans performed near the 1D
AF zone center for different transverse wave vectors. (c)
Constant-energy scans along the $l$ direction measured just above
the gap energy. Solid lines are fits to the data as described in
the text.} \label{tas}
\end{figure}

\begin{figure}[btp]
\includegraphics[width=3.2in]{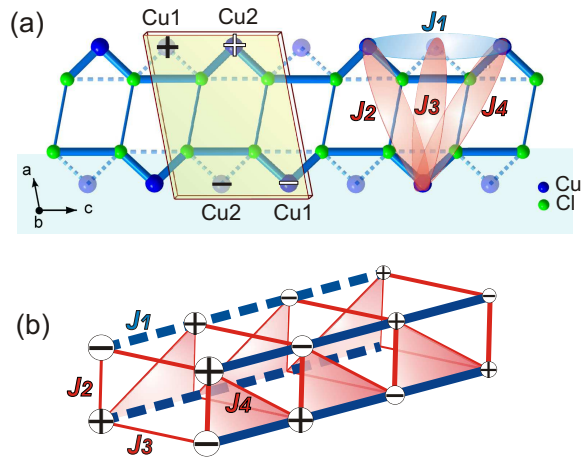}
\caption{(a) Ladder-like structure consisting of Cu$^{2+}$ ions
coupled through double chlorine bridges in \Sul. (b) A
topologically equivalent uniform 4-leg spin tube structure. The
relative instantaneous orientations of the spins are indicated by
``+'' and ``-'' signs.} \label{models}
\end{figure}

A survey of the reciprocal space by a series of constant-$q$ and
constant-$E$ scans revealed the inadequacy of the original model
for \Sul. Figure \ref{tof} shows the energy-$l$ projection of the
TOF spectrum measured at $T=70$~mK. It is immediately apparent
that the 1D AF zone-center, where the spin gap $\Delta$ is
observed as a minimum in the magnetic excitation spectrum, is
located near {\it half-integer} values of $l$. No low-energy
magnetic scattering was detected at integer $l$. This behavior
indicates that it is the next-nearest neighbor (NNN) spins in each
chain which are strongly antiferromagnetically correlated, not the
nearest-neighbor ones as proposed in Refs.~\cite{Fujisawa03,Fujisawa05}. The magnetic interactions
responsible for strong NNN in-chain AF coupling are readily
identified. They are established by superexchange across
two-chlorine bridges, such as Cu$_{1}$-Cl-Cl-Cu$_{1}$ or
Cu$_{2}$-Cl-Cl-Cu$_{2}$. As was previously found in the
structurally related material IPA-CuCl$_3$~\cite{Masuda06}, and in
the Ni-based compound NiCl$_2$-4SC(NH$_2$)$_2$ ~\cite{Zapf06},
Cu-Cl-Cl-Cu couplings can be much stronger than those across
single-halide Cu-Cl-Cu bridge due to more favorable bonding angles
and orbital overlap. Thus, each alternating spin chain in \Sul\
should actually be considered as a pair of superimposed AF spin
chains, each with a uniform (non-alternating) bond length equal to
$c$. Such spin chains are highlighted by bold or dashed
Cu-Cl-Cl-Cu bonds in Fig.~\ref{models}a.

\begin{figure}[tp]
\includegraphics[width=3.1in]{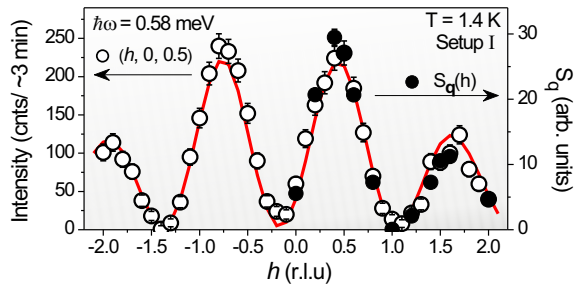}
\caption{Measured $h$-modulation of the scattering intensity of
the energy-gap mode at T = 1.4 K. Symbols and lines are as
described in the text.} \label{modulation}
\end{figure}

Our neutron data indicate that \Sul\ is indeed an exceptionally
one-dimensional magnetic material. The dispersion of magnetic
excitations is practically absent along the $a^\ast$ and $b^\ast$
axes (Fig.~\ref{tas}a,b), but very steep along the $c^\ast$
direction (Fig.~\ref{tof}). Only the highest-resolution Setup III
allowed us to estimate the corresponding spin wave velocity.
Typical constant-$E$ scans are shown in Fig.~\ref{tas}c. At $\hbar
\omega=0.6$~meV, just above the gap energy, resolution-limited
peaks are observed. However, at 2.5~meV energy transfer, an
intrinsic $l$-width becomes apparent and is due to a finite magnon
velocity. As will be discussed below, the peaks are actually
centered at {\it incommensurate} zone-centers $l_0=0.5-\delta$ and
$l_0=1.5+\delta$, $\delta=0.022(2)$. The approximate dispersion
relation for low-energy magnons is therefore written as:
 \begin{equation}
 (\hbar \omega_\mathbf{q})^2 = \Delta^2 + v^2(\vec{\mathbf{q}}\vec{\mathbf{c}}-2\pi l_0)^2. \label{disp}
 \end{equation}
A simple model cross section function can then be constructed
using the single-mode approximation:
\begin{equation}
 \frac{d^2 \sigma}{d \Omega d E'} \propto |f(\mathbf{q})|^2
 \frac{|F(\mathbf{q})|^2}{\hbar
 \omega_\mathbf{q}} \delta(\omega-\omega_\mathbf{q}), \label{sqw}
\end{equation}
where $F(\mathbf{q})$ is the {\it a priori} unknown structure
factor that is expected to be a smooth function of $\mathbf{q}$,
and $f(q)$ is the magnetic form factor for Cu$^{2+}$.
Equation~\ref{sqw} convoluted with the known spectrometer
resolution function was fit to the bulk of the 3-axis data to
determine $\Delta\simeq0.55$~meV and $v\simeq14$~meV. Excellent fits are
obtained and shown in solid lines in Fig.~\ref{tas}. The resulting
dispersion curve is plotted over the TOF data in Fig.~\ref{tof}.

Having determined that Cu-Cl-Cl-Cu pathways dominate in-chain
interactions, we now have to consider their role in the
inter-chain coupling. As illustrated in Fig.~\ref{structure}, such
bridges link pairs of adjacent Cu$^{2+}$ chains in a plane {\it
perpendicular} to the $b$ axis. Note that these are {\it not} the
same chain pairs as those considered in Ref.~\cite{Fujisawa03}.
Recalling that each structural Cu$^{2+}$ chain actually consists
of two superimposed Heisenberg spin chains formed by NNN AF
interactions, we find that Cu-Cl-Cl-Cu bridges in \Sul\ establish
4-leg spin ``ladders'', as shown in Fig.~\ref{models}a. In this
representation $J_1$ denotes the leg coupling responsible for a
strong $c$-axis dispersion of magnetic excitations. In addition,
three distinct ``rung'' interactions $J_2$, $J_3$ and $J_4$ can be
identified. None of these are affected by the bond alternation
between the two distinct sites, Cu1 and Cu2. The four exchange
pathways give \Sul\  the unique topology of a 4-leg spin tube, as
shown in Fig.~\ref{models}b.

The interactions $J_2$--$J_4$ establish correlations between the
individual legs of each spin tube. Due to the resulting
interference effect, the intensity of the gap mode becomes
dependent on momentum transfer perpendicular to the chain axis. A
constant-$E$ scan taken using Setup I just above the gap energy
along the $(h,0,0.5)$ direction, near the 1D zone-center, is shown
in Fig.~\ref{modulation} (open circles) and reveals a strong
intensity modulation. The solid symbols represent integrated
intensities measured in $l$-scans of the type shown in
Fig.~\ref{tas}c. The observed modulation is periodic, yet
off-centered relative to $h=0$. Similar behavior is seen at
$(h,0,1.5)$. To explain this phenomenon quantitatively, we recall
that in a simple 2-leg spin ladder the gap modes correspond to
spins fluctuating coherently on the two ladder
legs~\cite{Barnes93}. Their intensity is modulated by the
modulus-squared of the rung's structure factor. For AF rungs
defined by a vector $\mathbf{d}$, $|F(\mathbf{q})|^2=
\sin^2(\mathbf{q}\mathbf{d}/2)$. By analogy, the lowest-energy
excitation of the 4-leg spin tube in \Sul\ is a coherent
fluctuation of its four legs. The relative phases of the legs will
depend on the details of the rung coupling. The intensity of the
gap mode becomes modulated by the structure factor of the rung
unit comprised of four spins, one on each leg:
\begin{equation}
 F(\mathbf{q})=\sum_{j=1}^4\phi_j\exp(-i\mathbf{q}\mathbf{r}_j).
\end{equation}
Here $j$ labels the spins on the rung unit, and $\phi_j=\pm 1$ are
their corresponding phases. Since in our model inter-leg
correlations are established by $J_2$, $J_3$ and $J_4$, only those
spins directly connected by these interactions need be considered
as members of a rung unit. Having tried all possible site and
phase combinations, we found that only one can reproduce the
measured transverse intensity modulation. The thus identified
rung-unit is shown in Fig.~\ref{models}a and coincides with the
nominal crystallographic unit cell. The corresponding phases of
its four spins are indicated by ``+'' and ``-'' signs. This model,
with ionic magnetic form factors and resolution effects taken into
account, reproduces the observed intensity modulation almost
perfectly, as shown in a solid line in Fig.~\ref{modulation}. The
4-leg topology may also be responsible for the relatively small
excitation gap in \Sul\ \cite{White94,Frischmuth96}.

Our model naturally accounts for the observed shift of the 1D AF
zone-center relative to the commensurate position. We note that
the diagonal AF interaction $J_4$ incurs a geometric frustration
relative to $J_1$--$J_3$ (Fig.~\ref{models}b). In classical
magnets such frustration induces a helimagnetic state
\cite{Yoshimori}. Theory predicts that in a frustrated quantum
2-leg spin ladder this incommensurability may survive in the
equal-time spin correlation function \cite{White96}, even though the
ground state is disordered. In this case the minimum of the magnon
dispersion occurs at an incommensurate wave vector. The situation
in  \Sul\ is a direct analogue of this effect in a spin tube with
frustrating diagonals. Such a singlet ground state with
incommensurate correlations has not been found in any previously
studied 1D spin gap system.

We conclude that \Sul\ realizes the rare 4-leg spin tube model. The mapping is actually quite accurate: it is not at all perturbed by the intrinsic structural modulation of the Cu$^{2+}$ chains, nor by three-dimensional interactions that are negligible. The spin tube Hamiltonian posses unique quantum mechanical properties, due to periodic boundary conditions along the inter-chain direction. However, for previous lack of prototype materials, only a limited number of theoretical results have been accumulated~\cite{Sato07,Schnack,Fouet06}. Future work on \Sul\ should focus on higher-energy excitations, where one can expect to find exotic states with higher values of the transverse discrete quantum number. Magnetic field experiments will probe Bose-Einstein condensation of magnons. Due to geometric frustration, one can expect a unique helimagnetic high-field ordered phase.

Research at ORNL was funded by the United States Department of
Energy, Office of Basic Energy Sciences- Materials Science, under
Contract No. DE-AC05-00OR22725 with UT-Battelle, LLC. The work at
NCNR was supported by the National Science Foundation under
Agreement Nos. DMR-9986442, -0086210, and -0454672. The authors
are grateful to Drs. B. Sales and D. Mandrus for making available
their laboratory for the sample preparation. We also thank Drs. K.
Hradil, P. B\"{o}ni and R. Mole for their help in the preliminary
INS experiments on the 3-axis spectrometer PUMA, at FRM II,
Germany.

\end{document}